\documentclass[twoside,onecolumn,journal]{CBMStran}

\usepackage{amsfonts,amssymb,amsmath,amsthm}
\usepackage{blindtext}
\usepackage{graphicx}
\usepackage{epstopdf} 
\usepackage{textcomp}
\usepackage{lipsum}
\usepackage[bottom=3cm,top=4cm,left=3cm,right=3cm]{geometry}
\usepackage{fancyhdr}
\usepackage{lastpage}
\usepackage{booktabs}
\usepackage{threeparttable}
\usepackage[font=small]{caption}
\usepackage[colorlinks=true]{hyperref}
\usepackage{xcolor}
\usepackage{titlesec}
\hypersetup{linkcolor=cyan, citecolor=green, urlcolor=blue}
\makeatletter
\renewcommand*{\@seccntformat}[1]{\csname the#1\endcsname\hspace{0.2cm}}
\makeatother

\renewcommand{\figurename}{Figure}
\renewcommand{\tablename}{Table}
\renewcommand{\thetable}{\arabic{table}}
 
\begin{document}
\title{\Large \textbf{The interplay of common noise and finite pulses on biological neurons}}

%
%
\author{%
	{\textbf{\normalsize Afifurrahman$^{1*}\footnote{*Corresponding author}$, 
	Mohd Hafiz Mohd$^{2}$, Farah Aini 
	Abdullah$^{3}$}}\\
	\vspace*{0.3cm}
	\small ${^1}$Department of Mathematics, Universitas Islam Negeri Mataram, Mataram 83116, Indonesia\\
	$^{2,3}$School of Mathematical Sciences, Universiti Sains Malaysia, 11800 USM, Penang, Malaysia\\

	\vspace*{0.2cm}
	*Email: afif.rahman@uinmataram.ac.id 
	\thanks{\noindent Received (date), Revised (date), Accepted for publication (date). Copyright \textcopyright 2020 Published by Indonesian Biomathematical Society, e-ISSN: 2549-2896, DOI:10.5614/cbms.2020.1.1.(Paper's number)}}



%
%

\markboth{COMMUN. BIOMATH. SCI., Vol. (), No. (),  (year), pp. (first page-last page)}
{}
%



\maketitle

\begin{abstract}
	
	The response of neurons is highly sensitive to the stimulus. The stimulus can be associated
with a direct injection in {vitro} experimentation (e.g., time dependent and independent
inputs); or post-synaptic potentials resulting from the interaction of many neurons.
{A typical incoming stimulus resembles a noise which in principle can
be described as a random variable. In computational neuroscience, the noise has been
extensively studied for different setups}. In this study, we
investigate the effect of noisy inputs in a minimal network of 
two {identical} leaky integrate-and-fire (LIF) neurons
interacting {with finite pulses}. {In particular,} we consider a Gaussian white
noise as a standard function for stochastic modelling of neurons, 
while taking into account the pulse width as
an elementary component {for the signal
transmission}. By exploring the role of noise and finite pulses, the two neurons show a synchronous spiking behaviour characterized by fluctuations in the inter-spike intervals. 
Above some critical values the synchronous regime 
collapses onto
asynchronous dynamics. The abrupt change in such dynamics is
accompanied by a hysteresis, i.e., the coexistence of synchronous and asynchronous firing
behaviour.
	
	\vspace*{0.3cm}
	\noindent \textit{Keywords: LIF neuron, stochastic process, finite pulses, inter-spike interval, synchronization.}
	
	\vspace*{0.1cm}
	\noindent \textit{2010 MSC classification number: 37N25, 92B25, 92C42}
\end{abstract}
%


%
\CBMSpeerreviewmaketitle

\section{Introduction}
\label{intro} 

A single neuron reacts to the incoming stimuli in unexpected ways. {As for instance, 
it might fires a pulse (spike) at constant rates, 
if the neuron operates above a threshold. The periodic firing 
behaviour is probably the simplest dynamical phenomena observed in the neuronal models  \cite{Beeman2014,Gerstner2014,Sherwood2014,Liu2014}.}

{The variability of neuron's response to the incoming stimulus} is encoded in the so-called inter-spike interval (ISI), {i.e., the time-length between any two consecutive spikes 
\cite{Gerstein1960,Christodoulou2001}}. In a situation where the periodic regime appears, the coefficient of variations of ISIs is zero. Here, one can easily predict a sequence of spiking times once the neuron fires
its first spike.

Unfortunately, the response of real neurons is far from being regular and periodic. Instead, the irregular 
{firing activity} is the {most prominent feature 
and it might be related to brain disorders according to some experimental studies \cite{Barrett2017,
Rong2021,Wu2022}}. The irregularity is characterized by the variability {on} 
ISIs, which means that the fluctuation is not zero at all \cite{Taube2010}. Investigating the necessary conditions in which irregular (and regular) behaviours exist is upmost important as it allows us to understand how the brain perceives the world.

Multiple sources are mainly responsible for the emergence of irregular behaviour. In the cortical neurons, for example, the intrinsic noise resulting from the fast activation of ion channels may contribute to the variability {on ISIs} \cite{Stiefel2013}.  
{The noise has been extensively studied for different setups
and conditions. It is believed that the noise  
is not only slows down but also improves the reliability of neuronal firing, as well as facilitates the temporal patterning of synchronization and chaos
 \cite{Ermentrout2018,McDonnell2011,Shi2008,Zirkle2021,Peng2022,Leng2020}.}

Apart from the noise effect, the irregular {firing} behaviour is also induced by some chemical-type currents which highlight a significant impact of synaptic transmission 
\cite{Durstewitz2007}. {The real neurons communicate 
by sending and receiving pulse that lasts about 1-2 ms time course \cite{Gerstner2014}.
A recent study reports the pulse width
is responsible for the irregular response of neurons in auditory systems \cite{Zhang2022}.
 } 
 It {also} has been suggested that the persistence of irregular dynamics in 
{coupled oscillators} is strongly depends on the duration of the pulse
\cite{Afif2021,pietras2023} {making it plausible for neuronal modelling}. 

{In this paper, we investigate numerically
the stochastic neural networks taking into
account the pulse width as an additional parameterisation for 
the synaptic coupling, which has never been considered 
in any other similar works (see \cite{Mishra2005,Shi2008,Zirkle2021}).}
{Moreover, compared to the previous studies, our object of study is a one-dimensional leaky integrate-and-fire (LIF) neuron \cite{Burkitt2006I,Burkitt2006II}.
The model is quite simple yet the dynamics are very rich.

On top of that, the earlier works on stochastic LIF model were mostly dealing 
with a single
neuron and concern only to the 
theoretical perspectives of it while less attention 
has been given to the emerging dynamical
states in macroscopic scales \cite{Ostojic2011,Dumont2016,Pirozzi2018,Thieu2022}.
Here,
We are working with a network of two noisy LIF neurons interacting via excitatory synapses
to test the robustness of irregular firing activities.}
The synaptic transmission amongst them is modelled as a smooth function of exponential shape. Due to the randomness of stochastic input, we limit ourselves to a numerical analysis.

More precisely, in Sec. \ref{s:2} we define the model including the Gaussian white noise and finite width pulse. In the same section, we introduce the schemes used to perform numerical simulation and the corresponding statistical quantities, namely the synchrony measure, to assess the emerging dynamical regimes. In the Sec. 
\ref{s:3} we present some findings regarding the effect of noise on single neuron, 
which later on generalized for the two neurons interacting through exponential pulses. Finally, in Section \ref{s:4} we discuss the limitations of this study and highlight some problems for the future.


\section{Model formulation}\label{s:2}

\begin{figure}[ht]
\centering
\includegraphics[width=8.8cm]{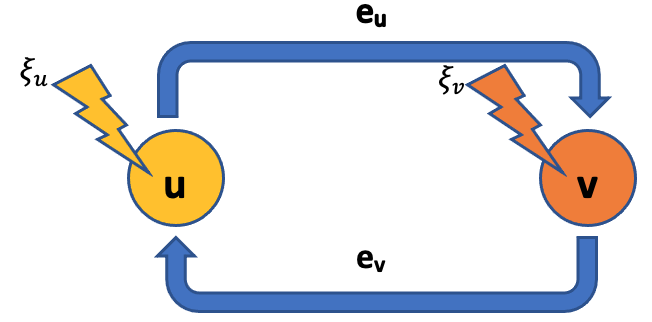}
\caption{A schematic representation for model \eqref{eq:1}}\label{fig:1}
\end{figure}

The starting point of this paper is a {general} stochastic system (Equation 2) introduced in \cite{Shi2008}. 
Instead of electrical coupling, we incorporate the synaptic type into the model by employing the finite width pulses. The unit cell is modelled as a LIF neuron defined in
\cite{Politi2010}. 
{We use a simple network architecture
of two cells interacting via excitatory synapse with the same coupling strength
as studied in \cite{Zirkle2021}.}
More specifically, our systems (see Figure \ref{fig:1}) consist of two {identical and} mutually coupled LIF neurons given by the following differential equations:
\begin{equation}\label{eq:1}
   \dot{u}=a-u+\xi_u(t)+\frac{\mu}{2}e_u,\qquad \dot{v}=a-v+\xi_v(t)+\frac{\mu}{2}e_v,
\end{equation}
{combined with the resetting rules: if $u\geq u_{th} \Rightarrow u=u_r$
and $v\geq v_{th} \Rightarrow v=v_r$, respectively.}
{The variables} 
$u$ and $v$ represent the membrane potentials for each neuron, {$\dot{u}$ and 
$\dot{v}$ 
are the corresponding derivatives with respect to time $t$},
while $a$ is constant current. We assume that all variables and parameters are dimensionless. The parameter $\mu>0$ is the coupling strength describing the intensity of excitatory postsynaptic potential. We introduce a scaling factor of $1/2$ to account for the full connectivity in the present model of two interconnected neurons. In principle, the model can be generalized to $N$ number of fully coupled oscillators, so the appropriate scaling would be $1/N$ \cite{Gerstner2014}. 

Following ref \cite{Afif2021,Afif2020}, the fields $e_u$ and $e_v$ are the linear superpositions of exponential spikes emitted by the sending neurons. Mathematically,
\begin{equation}\label{eq:2}
   \dot{e}_u=-\alpha\left(e_u - \sum_{k} 
   \delta \left( t - t^{v}_{k}\right) \right),\qquad 
   \dot{e}_v=-\alpha\left(e_v - \sum_{k} \delta \left( t - t^{u}_{k}\right) \right),
\end{equation}
where $\alpha$ denotes the inverse width of the incoming excitatory pulses, and both 
$t^{u}_{k}, t^{v}_{k}$ are the emission times of the {$k$}th 
pulse sent by the neuron 
$u$ and $v$, respectively. Whenever $u,v$ reaches the threshold $u_{th}, v_{th}$, then the two neurons are reset to $u_r,v_r$. 
In the limit of $\alpha\rightarrow\infty$, the Equations \eqref{eq:2} 
are simply the sum of $\delta$ pulses, i.e.,
\begin{equation}\label{eq:3}
   e_u = \sum_{k} \delta \left( t - t^{v}_{k}\right),\qquad 
   e_v = \sum_{k} \delta \left( t - t^{u}_{k}\right).
\end{equation}
The functions $\xi_u(t)$ and $\xi_v(t)$ are Gaussian white noise with the statistical property: 
$\mathbf{E}\left[\xi(t)\right]=0$ for all time $t$, and the autocorrelations 
$\mathbf{E}\left[\xi(t')\xi(t'')\right]=\sigma^2\delta\left(t'-t''\right)$ for $t'\neq t''$, 
{where $\mathbf{E}$ denotes the expected value}. Here, $\sigma$  is the noise amplitude and $\delta (t)$ is the delta function. 
For simplicity, the functions $\xi_u (t)=\xi_v (t)=\xi (t)$, which assumes the same noisy inputs for the two neurons.

\subsection{Numerical analysis}\label{s:2.1}
The Euler-Maruyama scheme \cite{Bayram2018} is implemented to {solve}
simultaneously the main Equations \eqref{eq:1}-\eqref{eq:2}.
Notice that, as the integration 
time step $dt\rightarrow 0$ the trajectory is smoothened since the fluctuations 
term vanishes with $dt$. For convenience, we choose $dt=10^{-3}$.

The initial conditions for the membrane potential are drawn randomly from a uniform distribution in a closed interval $[0,\epsilon]$  i.e., $(u(0),v(0))\in U_{[0,\epsilon]}$  while the fields are initially set to 0. Table 1 summarizes all parameters being used for the numerical simulation.
{Both reset potential $(u_r,v_r)$ and threshold $(u_{th},v_{th})$ have been chosen
following the hypothetical values in the deterministic version of LIF neurons \cite{Politi2010}. 
The spike emission of LIF neuron is 
achieved when the constant current is larger than the threshold, which in this case: $a>1$. Here we choose $a=1.5$ as
initially set in \cite{Gerstner2014}. 
We also assume a relatively weak coupling strength in the order of magnitude $10^{-3}$
as introduced in
\cite{Zirkle2021}.  Finally, the noise amplitudes and the pulse widths have been
selected according to the computational experimentations \cite{Afif2021,Shi2008,Dumont2016}.}

\begin{table}
\centering
\begin{tabular}{ |c|c|c| } 
 \hline
 Parameters                               &                    & References \\
 \hline\hline 
 Threshold: $u_{th},v_{th}$        & 1                 & \cite{Politi2010} \\ 
 \hline
 Reset potential: $u_r,v_r$        & 0                 & \cite{Politi2010} \\ 
 \hline
 Constant current: $a$          & 1.5              & \cite{Gerstner2014} \\
 \hline
 Coupling strength: $\mu$   & $2\times 10^{-3}$		& \cite{Zirkle2021} \\
 \hline
  Noise amplitude: $\sigma$     & $[0.1, 1.4]$ & \cite{Shi2008} \\
 \hline
 Inverse pulse width: $\alpha$  & $[20, 95]$  & \cite{Afif2021} \\
 
 \hline
\end{tabular}
\caption{The default parameters used during the simulation 
{for Eqs.\eqref{eq:1}-\eqref{eq:2}.}}\label{tab:1}
\end{table}

{
In order to quantify the degree of synchronization in coupled 
systems \eqref{eq:1}-\eqref{eq:2},
we make use of the average technique \cite{Shi2008}
 \begin{equation}\label{eq:8}
    R = \langle \sqrt{\left(v-u\right)^2 + \left(e_v - e_u\right)^2} \rangle.
 \end{equation}
The number $R$ is a synchronization error and the angular 
bracket $\langle \star \rangle$ denotes time averaging. 
If the two neurons are completely synchronized then, $R=0$.
Complete synchronization refers to a dynamical regime where all neurons within the network
are firing exactly at the same time.

}


\section{Main Results}\label{s:3}

\subsection{Single neuron}\label{s:3.1}

{
In the absence of any synaptic interaction ($\mu=0$), the two neurons are indeed identical
and evolve independently in time. 
In this case, the fields \eqref{eq:2} become
irrelevant and hence the problem of coupled systems \eqref{eq:1} can be reduced 
to a single neuron.
The time evolution for a single neuron is governed by the following stochastic system:
\begin{equation}\label{eq:4}
   \dot{y}=a-y+\xi (t), \hspace{0.2cm} {\text{if}} \hspace{0.1cm} y\geq 1 \Rightarrow y = 0,
\end{equation}
where $y$ is the membrane potential.
The Equation \eqref{eq:4} is a special case of the well-known Ornstein-Uhlenbeck process.
To simulate the trajectory of stochastic LIF neuron \eqref{eq:4}, 
it is reformulated as follows
\begin{equation}\label{eq:5}
   dy = \left(a - y\right)dt + \sigma\sqrt{dt}x
\end{equation}
where $x$ is a random number, drawn from a zero-mean Gaussian distribution with the unit variance. 
The same procedure of numerical integration is also adjusted 
for Equation \eqref{eq:5} as in the Section \ref{s:2.1}.}

The firing rate and coefficient of variations are statistical quantities 
frequently used to characterize the spiking behaviour
for single neuron 
\cite{Gerstner2014,Gabbiani2010}.
{The firing rate refers to the number of spikes divided by the length of integration time.
The explicit formula is given by }
 \begin{equation}\label{eq:6}
     \nu = \lim_{t\rightarrow\infty} \frac{n_t}{t}
 \end{equation}
 {where $n_t$ is the number of spikes emitted by the neuron
 over a time interval $t$.}

The coefficient of variations ($CV$) is a microscopic measure of the dynamics based 
on the variability of inter-spike intervals. 
{The inter-spike interval is 
time difference of two consecutive spikes, i.e., 
ISI$=t_{k+1} - t_k$.}
The $CV$ of ISIs is estimated by a ratio 
 \begin{equation}\label{eq:7}
     CV = \frac{\Delta}{\eta}
 \end{equation}
where $\Delta$ is standard deviation of ISIs and $\eta$ is the corresponding mean 
ISIs.

\begin{figure}[ht]
\centering
\includegraphics[width=12cm]{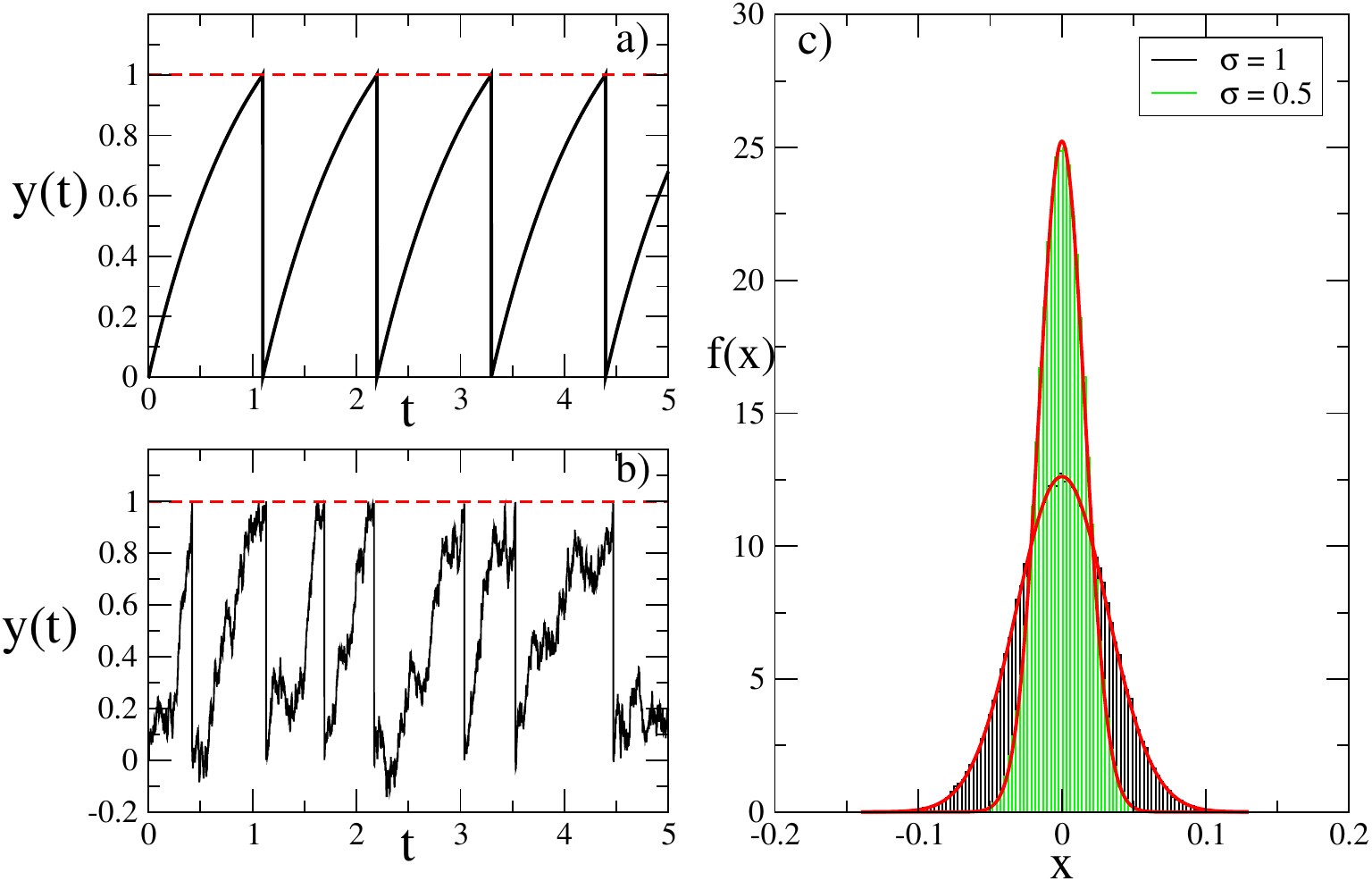}
\caption{(a)-(b) Time series of noise-free {($\sigma=0$)} 
and noisy trajectory {($\sigma=0.5$)} 
for LIF neuron 
{(see Eqs. \eqref{eq:4}-\eqref{eq:5}) with a fixed
 $a=1.5$}. 
 {When the membrane potential $y(t)$ hits 
 the threshold 1 (red dashed line), it is reset to 0.}
 (c) The probability density for white noise when $\sigma=0.5$ (green) 
and $\sigma=1$ 
(black) generated from the simulation. The solid red curves correspond to $f(x)$ in the Equation \ref{eq:10}. }\label{fig:2}
\end{figure}

When $\sigma=0$, {meaning that
no external noise is injected to the neuron}, the Equation \ref{eq:5} is just a linear 
{deterministic system}: 
\begin{equation}\label{eq:9}
{
d{y}=(a-y)dt. 
}
\end{equation}
The only relevant parameter here is $a$. 
{Note that, the neuron emits a spike if $a$ is larger than threshold 1, 
otherwise there is no spike emitted.} 
{The neuron's} free-noise trajectory is obtained by 
integrating Eq. \eqref{eq:9} with an initial condition $y(0)=0$ 
as shown in {Figure 
\ref{fig:2}a. Once the membrane potential passes a threshold of 1, it is reset to 0 while at 
the same time, the spike is produced.} 

The periodic firing behaviour of the neuron implies that $t_1$ 
defines a period of oscillation or ISI (for a detailed discussion on how to 
derive $t_1$ {explicitly} see \cite{Afif2023}). The next spikes consecutively occur at $ t_2=2t_1,t_3=3t_1,…t_i=it_1$ where $i$ is a positive integer.

Now, let us consider the dynamics of noisy neuron utilizing {$\sigma\neq 0$}. 
The noise is characterized by a probability density function 
\begin{equation}\label{eq:10}
f(x)=\frac{1}{\sigma\sqrt{2\pi dt}}\exp{\left(-\frac{1}{2dt}\left(\frac{x}{\sigma}\right)^2\right)}.
\end{equation}
Since the neuron is stimulated by the noise current with {an amplitude} $\sigma=0.5$ 
({green colour in} Figure \ref{fig:2}c), the inter-spike interval fluctuates irregularly 
in time (Figure \ref{fig:2}b). We can no longer predict the exact timing of the spikes because 
{the spike arrivals are independent of one another}. 
However, we may characterize the distribution of the spikes by 
splitting the length of integration time $t$ into $K$ bins such that
\begin{equation}\label{eq:11}
\tau = \frac{t}{K}.
\end{equation}
If $K$ is large (meaning that $\tau$ becomes sufficiently small)\footnote{This condition implies that any two or more spikes cannot occur together, instead, either exactly there is a single spike or no spike at all.}, the spike count in each bin is apt to be either 0 or 1. As a result, the distribution of the spikes is characterized by
\begin{equation}\label{eq:12}
   \mathbf{E}\left(\{0,1\}\right)=\frac{n_t}{K}=\nu\tau,\qquad 
   \mathbf{V}\left(\{0,1\}\right)=\nu\tau-(\nu\tau)^2,
\end{equation}
where $\mathbf{E}$ and $\mathbf{V}$ stand for the expectation and variance of the spike events, respectively. The second equality in the expected value results from the Equation \ref{eq:6} and \ref{eq:11}.

\begin{figure}[ht]
\centering
\includegraphics[width=6.cm]{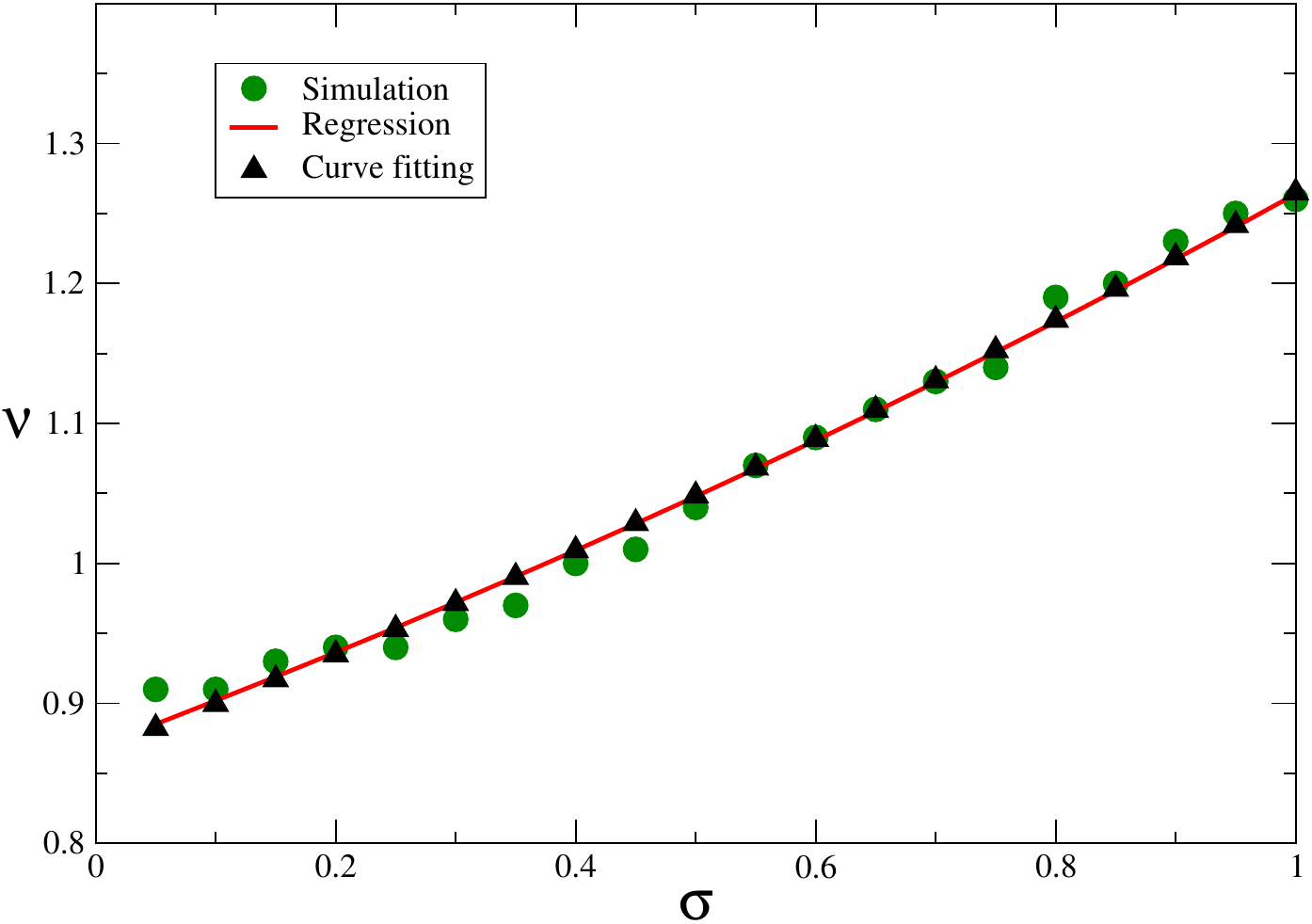}\quad
\includegraphics[width=6.2cm]{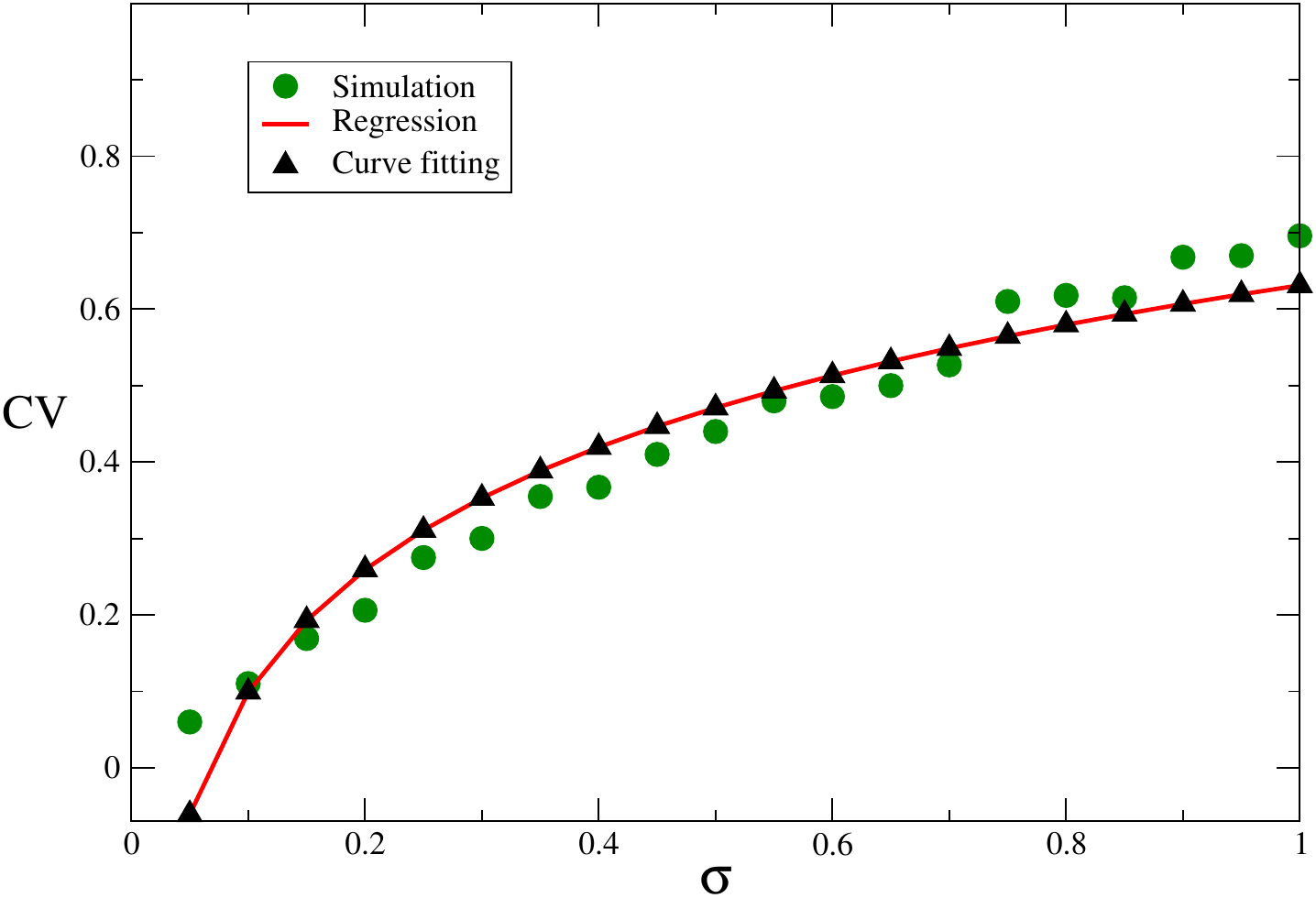}
\caption{Firing rate $\nu$ versus $\sigma$, coefficient of variations $CV$
versus $\sigma$.}\label{fig:3}
\end{figure}

Figure 
\ref{fig:3} shows a more quantitative description of neural activity. There one can see, the firing rate $\nu$ increases with the noise amplitude $\sigma$ meaning that the neuron is highly sensitive to the random stimulus. Such observation is in line with the experimental study, particularly for the neurons in the cerebral cortex \cite{Chaplin2017}. 

The same scenario is also shown when looking at the plot for the coefficient of variations $CV$ as a function of $\sigma$. The fluctuation of the inter-spike intervals is an effect of noisy inputs that raises by increasing the parameter $\sigma$. Such dynamics is robust, not only in the case of LIF neuron but also found in the biophysically Hodgkin-Huxley neuronal model \cite{Protachevicz2022}. The noise-induced spike {times} variability {serves as a 
basic mechanism for} the brain functioning such as information processing and perception \cite{Faisal2008}.

\subsection{Noise-induced synchronization {in two connected neurons}}\label{s:3.2}

We start our analysis of the noise effect on {
a network of two neurons ($\mu>0$)} interacting through finite width pulses.  
All parameters are kept the same as 
in Section \ref{s:2.1} with the initial conditions are drawn randomly in a such way that $(u(0),v(0)) \in U_{[0,1]}$ . The model \eqref{eq:1} assumes a weak coupling and so we fix the coupling strength as $\mu=2\times 10^{-3}$. The only relevant parameters are $\sigma$ and $\alpha$. The simulation is performed over 10000-time units after discarding 2000-time units of transients to let enough of the systems settle onto the attractor.

\begin{figure}[ht]
\centering
\includegraphics[width=9cm]{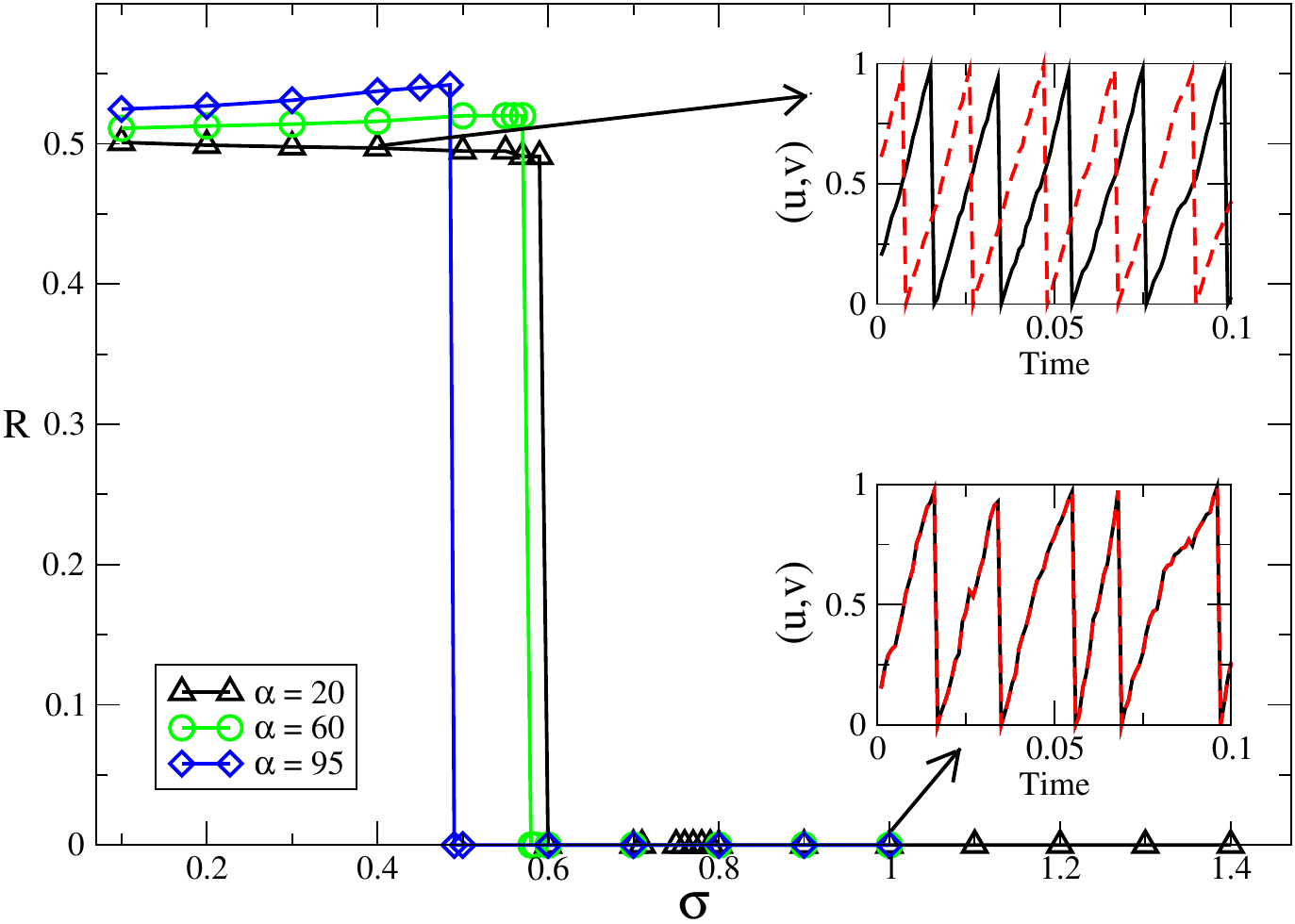}
\caption{Synchronization error $R$ as a function of noise amplitude $\sigma$ for different pulse widths $\alpha^{-1}$ : black triangles ($\alpha=20$), green circles ($\alpha=60$), and blue diamonds ($\alpha=95$). The two panels inset are the time series of the membrane potentials for $\sigma=0.4$ and $\sigma=1$, respectively, when $\alpha=20$. }\label{fig:4}
\end{figure}

We turn our attention on Figure \ref{fig:4} where we have plotted the synchrony measure upon varying the parameter $\sigma$ for a fixed $\alpha$. If $\alpha=20$, the synchronization error is approximately $R\approx0.5$ while the noise amplitude is within the interval range $\sigma\in [0.1, 0.59)$. In this regime, the two neurons are asynchronously firing the spike as shown by the time series of membrane potential $u$ (solid black line) and $v$ (red dashed line): see the upper panel inset generated for $\sigma=0.4$. 

Around $\sigma\approx0.6$ the synchronization error drops abruptly to 0 meaning that the two neurons are now completely synchronized. The time series analysis reveals that the ISIs fluctuate due to the presence of noise (lower panel inset).
For future reference, we define $\sigma_{\tau}$ as a transition point at which the systems collapse onto complete synchronization (CS).  Formally, 

\begin{equation}\label{eq:13}
   \sigma_{\tau} = \min \{\sigma | R\left(\sigma\right)=0 \}.
\end{equation}

It is also helpful to define a subset $S_{\tau}\subset S$
\begin{equation}\label{eq:14}
   S_{\tau} := \{\left(\alpha, \sigma_{\tau} \right) | R\left(\alpha, \sigma_{\tau} \right)=0 \},
\end{equation}
where $S := \{\left(\alpha, \sigma \right) | R\left(\alpha, \sigma \right)=0 \}$.
As instance, 
we observe the CS regime is self-sustained for $\alpha=20$ within the range of $\sigma\in [0.6, 1.4]$, where $\sigma_{\tau}\approx 0.6$ and $S_{\tau}=\{(20,0.6)\}$. Specifically, the time series of membrane potential show that $u(t)=v(t)$ for each $t$ and $t\rightarrow\infty$: see the lower panel inset generated for $\sigma=1$.

\begin{figure}
\centering
\includegraphics[width=9cm]{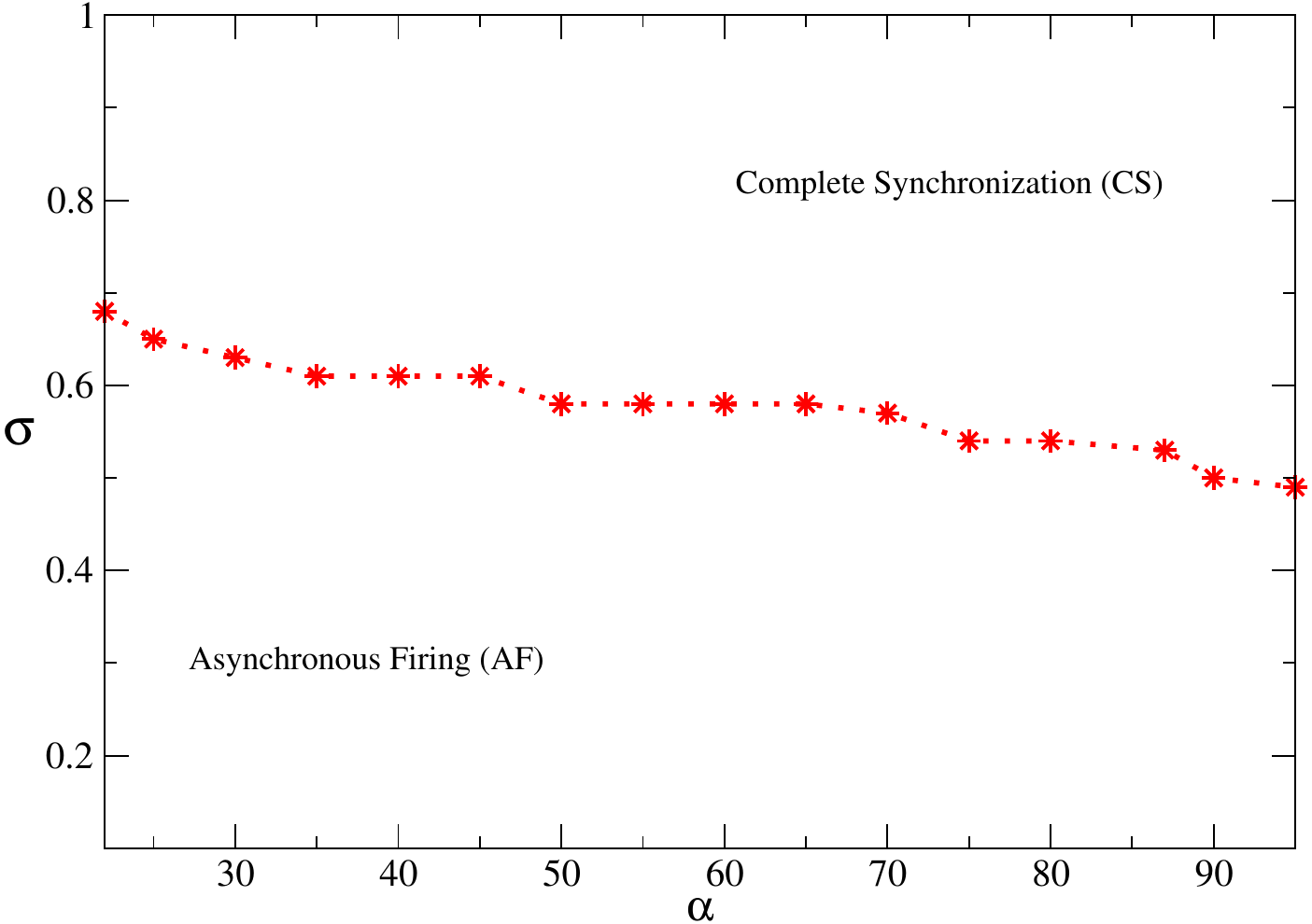}
\caption{Parameter space $\left(\alpha,\sigma\right)$. The red stars correspond 
to $S_{\tau}$.}\label{fig:5}
\end{figure}

Upon increasing $\alpha$ to 60 and 95, the asynchronous firing (AF) behaviour keeps persist in a finite but smaller interval range $\sigma\in [0.1, 0.58)$ and 
$\sigma \in [0.1, 0.49)$, respectively. Meanwhile, the CS emerges in a wider interval range $\sigma\in [0.58, 1.4]$ and $\sigma\in [0.49, 1.4]$.
 The Figure \ref{fig:5} sketches $S_{\tau}$ for $\alpha\in [22, 95]$ which are estimated for very long integration time. The curve $S_{\tau}$ splits the whole plane into two regions: complete synchronization (above the curve) and asynchronous dynamics (below one).

\begin{figure}[ht]
\centering
\includegraphics[width=10cm]{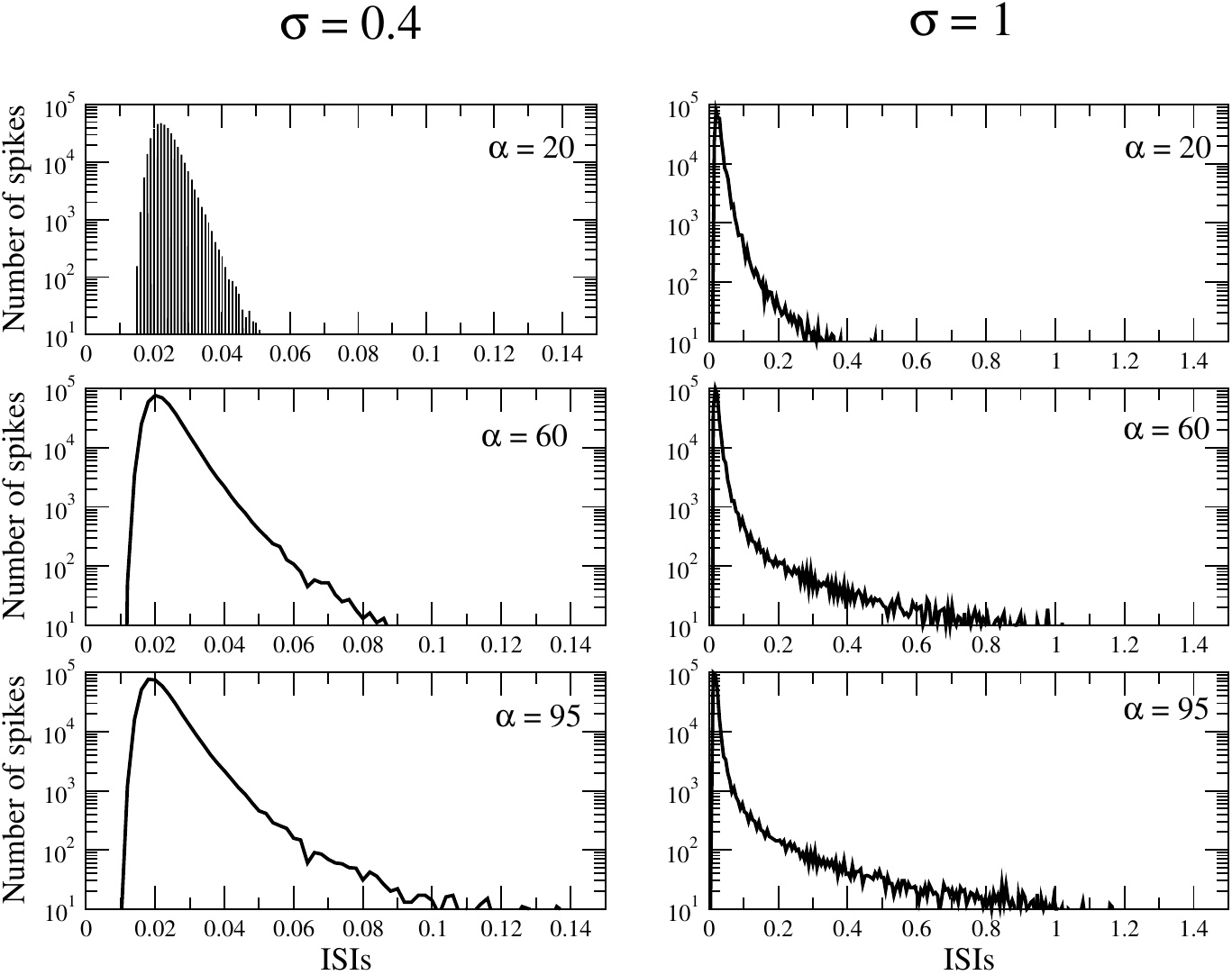}
\caption{Distribution of inter-spike intervals for $\sigma=0.4$ (left column) and 
$\sigma=1$ (right column). The vertical axis for all panels is in the logarithmic scale. Each row corresponds to different $\alpha$: upper panel ($\alpha=20$), middle panel ($\alpha=60$) and lower panel ($\alpha=95$).}\label{fig:6}
\end{figure}

To get a better insight into the role of noise and finite pulses to neuron dynamics, we produce a distribution of ISIs over 900000-time units for the neuron $u$ as displayed in Figure \ref{fig:6}. As seen, both noise and finite pulses give rise to the variations of ISIs. The width of the distributions is almost ten times wider as $\sigma$ is increased from 0.4 to 1 for each case of $\alpha$. Meanwhile, it is almost doubled upon shortening the pulse widths.

\subsection{Coexistence regime}\label{s:3.3}

So far, we have explored the behaviours of the coupled neurons (\ref{eq:1}) by selecting random initial conditions $(u(0),v(0))\in U_{[0,1]}$  and found a discontinuous transition upon tuning the parameter $\sigma$. Let us now investigate the possibility of coexistence regimes that are often accompanying the transition phenomena.

\begin{figure}[ht]
\centering
\includegraphics[width=10cm]{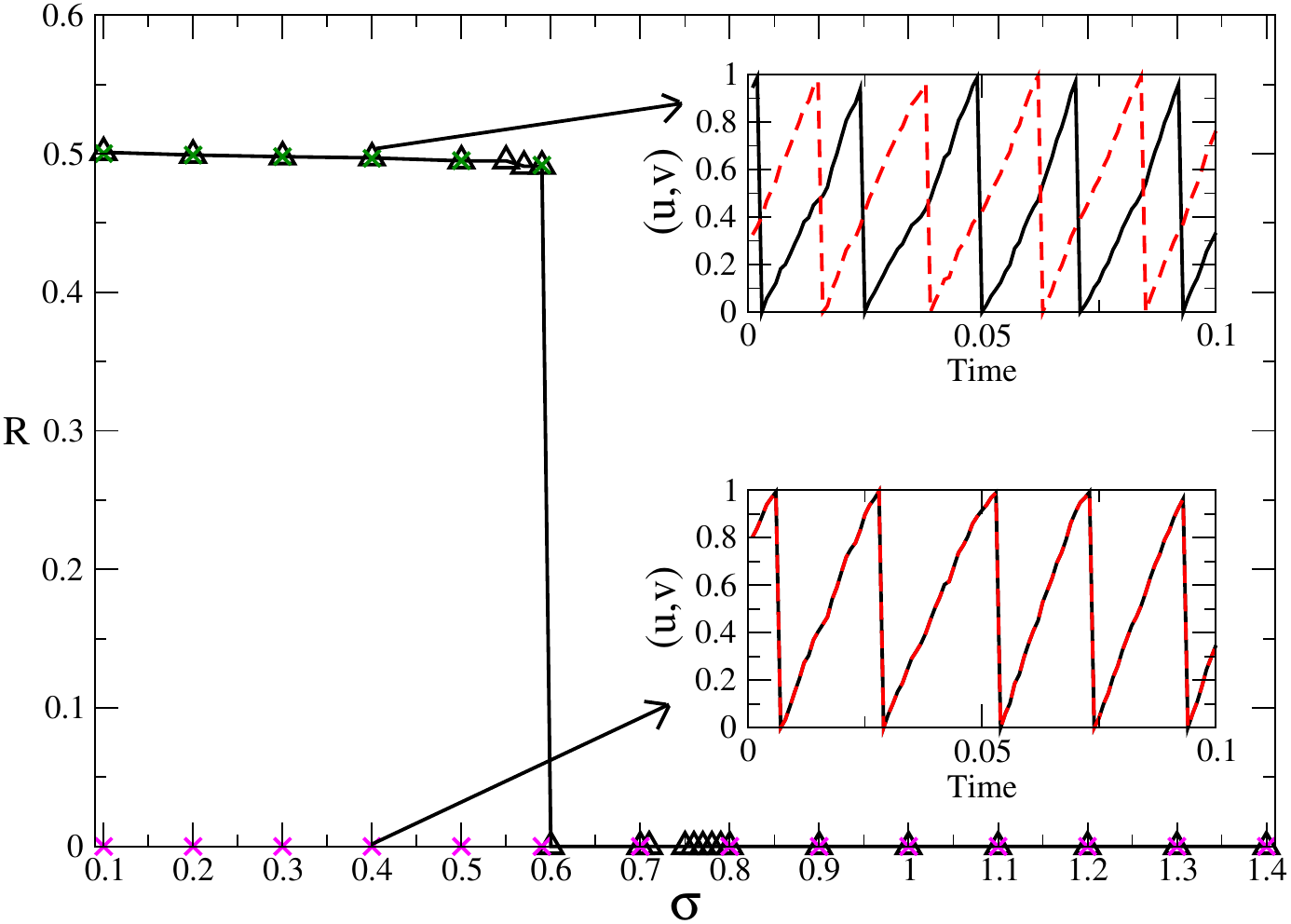}
\caption{The coexistence of complete synchronization and asynchronous firing dynamics for $\alpha=20$. The black triangles, magenta and green crosses correspond to different initial conditions: fully random (black triangles), finite but small interval width (magenta and green crosses). The narrow initial conditions are chosen to be in the order of $\epsilon=10^{-3}$ (magenta crosses) 
and $\epsilon=10^{-2}$ (green crosses). The time series corresponds to 
$\sigma=0.4$ with random initial conditions (upper panel) and narrow initial conditions (lower panel).}\label{fig:7}
\end{figure}

We start by fixing a small interval of width $\epsilon$ at which the initial conditions for neuron $u$ and $ v$ are selected whereas the fields are set to zero. Figure \ref{fig:7} compares the emerging scenario both for random and narrow initial conditions. In the interval range of  $\sigma\in [0.1, 0.6)$, the $R\approx 0.5$ results from imposing random initial conditions (see the black triangles). It is a clear indication of asynchronous firing dynamics. However, one can see that the complete synchronization emerges in the whole range of parameter $\sigma$ upon choosing a small
$\epsilon=10^{-3}$ as signalled by $R=0$ (see the magenta crosses). 

\begin{figure}
\centering
\includegraphics[width=6.cm]{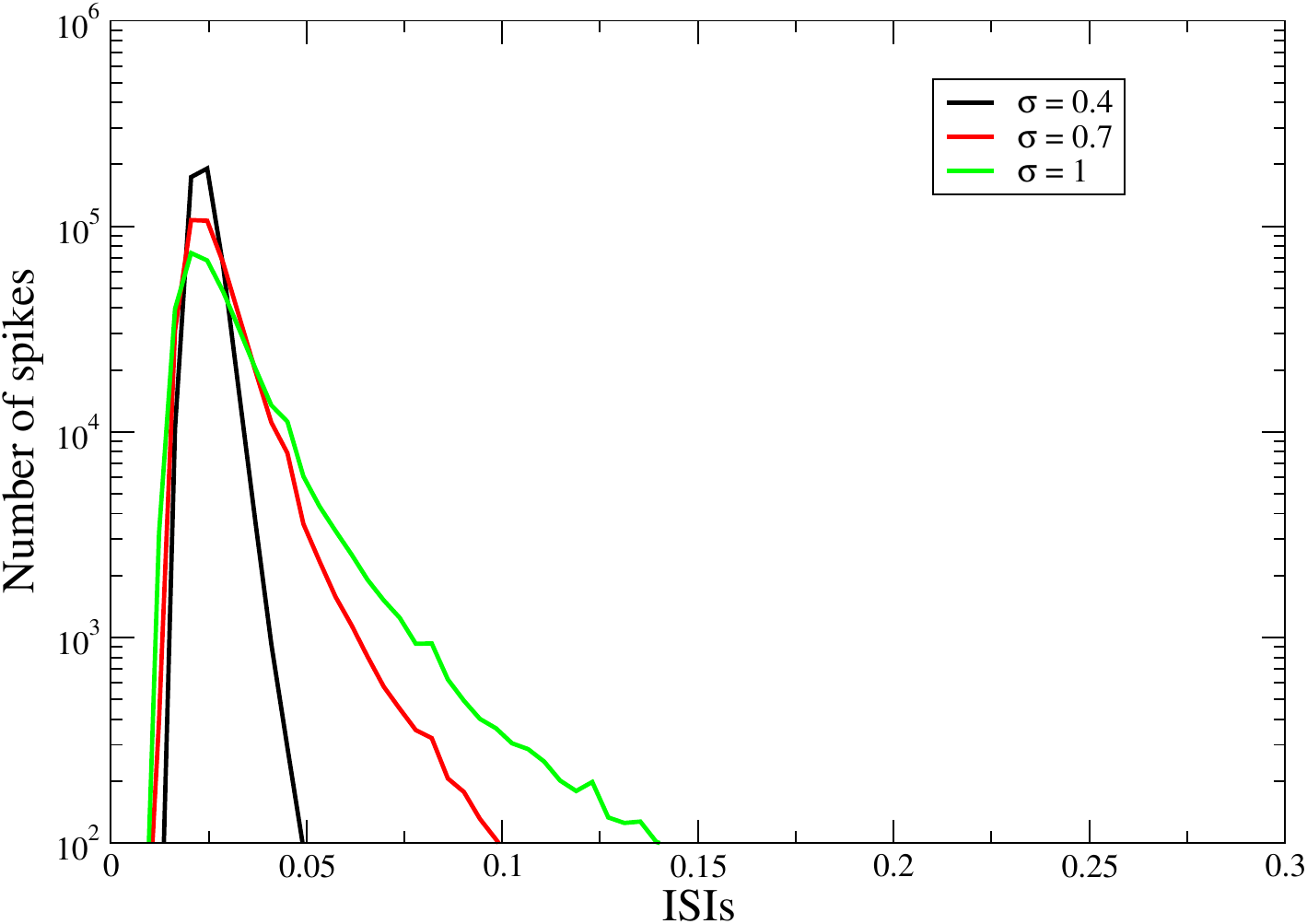}\quad
\includegraphics[width=6.cm]{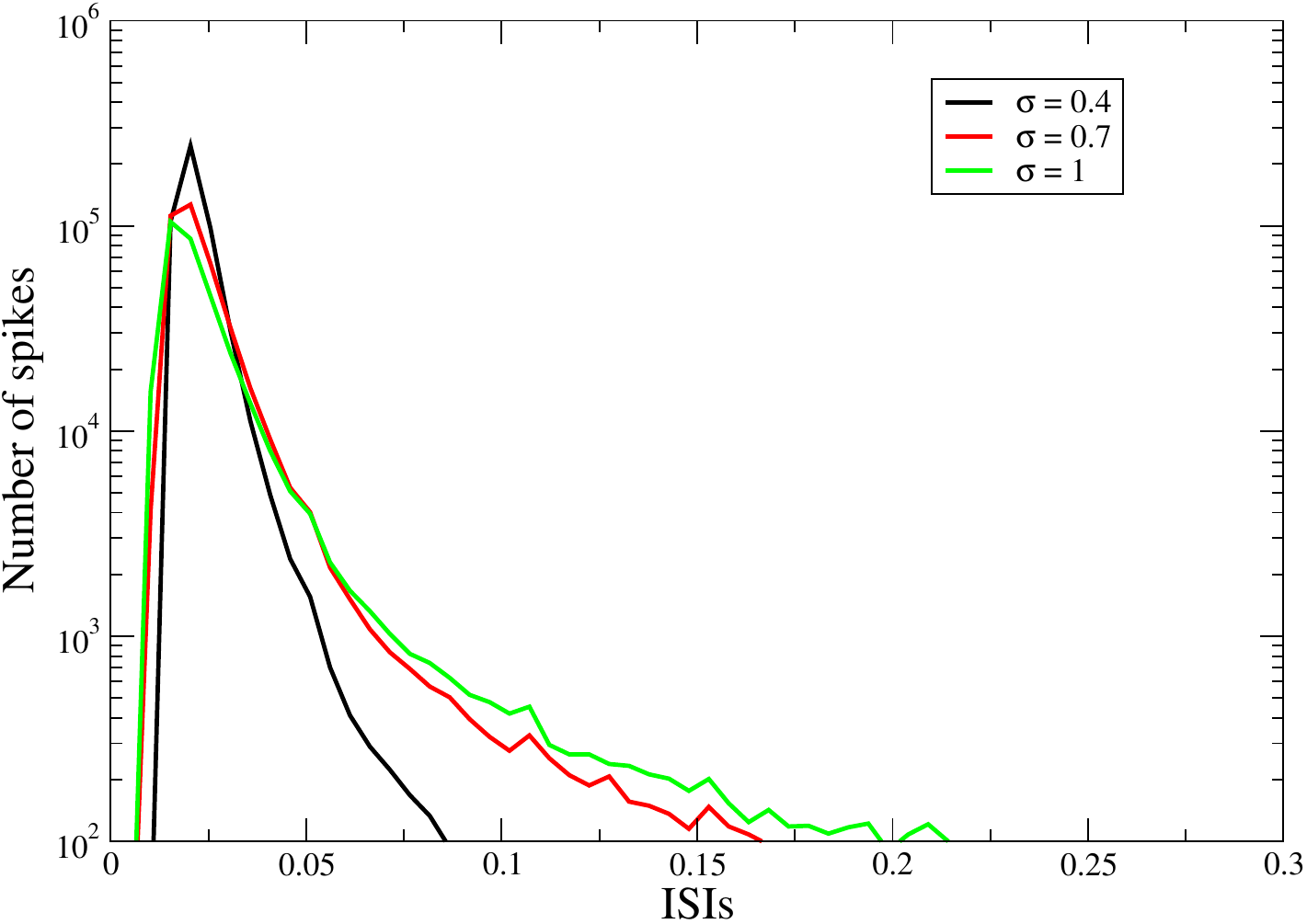}
\caption{Left to right: Distribution of inter-spike intervals for {neuron $u$ when} $\alpha=20$ 
and $\alpha=60 $. The  vertical axis is in the logarithmic scale. The colours are corresponding to different noise amplitudes: $\sigma=0.4$ (black), $\sigma=0.7$ (red), and 
$\sigma=1$ (green). The data are all generated for the narrow initial conditions $\epsilon=10^{-3}$.}\label{fig:8}
\end{figure}

The scenario is further supported by the time evolution of neuron $u$ (solid black) and $v$ (red dashed) for $\sigma=0.4$ in the inset panels. The upper panel correspond to the green crosses where $\epsilon \in (10^{-3}, 1]$ or equivalently random initial conditions. Clearly, the two neurons fire the spike asynchronously.  In comparison, the two neurons are identical for $\epsilon=10^{-3}$ as displayed in the lower panel.

Once again, the complete synchronous state is characterized by fluctuating ISIs. We sketch the distribution of ISIs {for neuron $u$}
 in Figure \ref{fig:8}. For a relatively low noise intensity (black curve), the ISIs looks homogeneous compared to the intermediate and strong (red and green curves) noises. Interestingly, decreasing the pulse width significantly adds the variability of ISIs. 

\section{Discussion and Conclusion}\label{s:4}
In this study, we have investigated the effect of {noisy inputs} on the firing patterns in a minimal network of two LIF neurons interacting through finite width pulses. Consistent to the earlier reports \cite{Shi2008,Zirkle2021,Kim2013}, the noise evokes the neuromodulation of synchronous and asynchronous firing dynamics. Some experimental studies show that both phenomena are robust in certain areas of the brain such as the visual cortex and hippocampus \cite{Chirwa1988,Gray1989}, which might be associated with brain disease such as seizures and epilepsy \cite{Scharfman2007}.

We observe the existence of synchronous behaviour, as indicated by $R=0$. The main important feature of the regime is the variability of ISIs, while at the same time the neurons are firing simultaneously. This study thereby suggests the dependence of the average ISI, not only on the coupling strength as highlighted in \cite{Kaltenbrunner2007}, but also on the noise amplitude. Given this situation, it is impossible to predict the exact timing of the spike compared to {the
 classical} 
case of “periodic” synchronization\footnote{“Periodic” synchronization refers to the regimes where the fluctuation of ISIs of the whole networks is zero on average \cite{Afif2021,Protachevicz2021}.}.  Consequently, the computation becomes more complicated as the noise interferes with the system: one must go beyond the traditional approaches at least in the context of stability analysis \cite{Laffargue2016}.
 
As we have shown, the fluctuation of ISIs changes when the noise is varied. A sufficient strong noise causes a large excursion on the trajectory and thus greatly affects the next spiking time once the last spike has fired. This is possibly the mechanism underlying the appearance of wide ISI distributions displayed in Figure 6 (e.g., for $\sigma=1$). Upon reducing the noise amplitudes, the two neurons desynchronize and fire the spike independently as signalled by $R>0$. In the microscopic level, the regime is characterized by more uniform ISIs and hence a shorter width of ISI distributions (Figure \ref{fig:6}, $\sigma=0.4$), similar to that have been observed in other neuronal models \cite{Mishra2005,Schwalger2010}.

The novel outcome of this study is the role played by the {width of} exponential pulses in shaping the ISI distribution. The width of ISI distribution is lengthened as the pulse {duration} is shortened (see again Figure \ref{fig:6} and \ref{fig:8}), which also suggesting the dependence of 
ISI {distribution} on the finite width pulses. It is unclear, however, to what extent the pulse width affects the distribution of ISIs produced by different levels of noisy inputs. This problem, in principle, can be approached mathematically by deriving an equation describing the probability distribution of ISIs by means of the Fokker-Planck equation \cite{Ostojic2011}, considering the pulse widths as the additional parameter.

Another observed phenomenon is the transition from complete synchronization to asynchronous dynamics. As indicated by the sharp jump of global observable variable $R$, the transition is discontinuous, a typical first-order transition commonly found in many living cells including neurons, as usual mechanism to respond to the external stimulus \cite{Fedosejevs2022}. 
In this context, the noise should be regarded as a crucial component in the mathematical modelling for the nervous systems. The real neurons are noisy, and the noise might come from many sources such as thermal noise, ion channel propagation, synaptic transmissions, {
network connectivity, etc \cite{Faisal2008,Serletis2011}}. Furthermore, the transition region is accompanied by a hysteresis. Upon choosing different initial conditions within a small interval range $\epsilon=10^{-3}$, we explore the emergence of bistable regimes: the new value of $R=0$ (see magenta crosses in Figure \ref{fig:7}) indicates a complete synchronous behaviour. 

{Our findings highlight the key points for future reference. First,
even in simple stochastic LIF neural networks, the dynamical regimes are intriguing and are in line with
the studies in a higher dimensional system, such as Hindmarsh-Rose and 
Morris-Lecar neurons (see again \cite{Shi2008,Zirkle2021}). 
We hypothesize the overall scenario emerging in this model is likely to persist in the larger system sizes. 
Secondly, the finite pulses is matter: shortening the pulse width adds the variability of
ISIs in stochastic neurons. If confirmed analytically, these statistics could lead
to well-established computational models and thus better understanding
of irregular activity in the brain.   
}

Finally, it is important to note some limitations of this study. We are working with a minimal network of two globally coupled {identical} neurons, while the brain consists of billions of neurons having more complex structures {and intrinsic properties}. 
{The works herein can be extended to large-scale networks
in the assumption of two interacting populations of excitatory and inhibitory neurons, which will
be our next object of study.}
This study is also limited to test numerically {the impacts of  noise and finite pulses on neuron dynamics (see again Figure \ref{fig:5})}. We argue that the asynchronous firing dynamics vanishes for a large $\alpha$, therefore further studies are required.

\section*{Acknowledgement}
This work was partially supported by the Indonesian Mathematical Society (IndoMS) through a visiting research program, grant No. 028/Pres/IndoMS/SP/VIII/2022.

\bibliographystyle{abbrv}

\end{document}